\documentclass{ws-procs9x6}

\usepackage{epsfig}

\newcommand{\lwig}{\mbox{\,\raisebox{.3ex}
    {$<$}$\!\!\!\!\!$\raisebox{-.9ex}{$\sim$}\,}}
\newcommand{\gwig}{\mbox{\,\raisebox{.3ex}
    {$>$}$\!\!\!\!\!$\raisebox{-.9ex}{$\sim$}}\,}
\newcommand{\lambdabar}{{\hbox{$\lambda_e$\kern-1.9ex\raise+0.45ex\hbox{--}
\kern+0.2ex}}}

\newif\ifhepph
\hepphtrue

\begin{document}

\title{
\ifhepph
\vspace{-3cm}
{\rm\normalsize\rightline{DESY 03-039}\rightline{\lowercase{hep-ph/0304139}}}
\vskip 1cm 
\fi
Boiling the Vacuum with an X-Ray Free Electron Laser\ifhepph\footnote{\uppercase{I}nvited talk 
presented at the
\uppercase{W}orkshop on \uppercase{QUANTUM ASPECTS OF BEAM PHY\-SICS}, 
\uppercase{J}anuary 7--11, 2003, \uppercase{H}iroshima, \uppercase{J}apan.}\fi
         }

\author{A.~Ringwald}

\address{Deutsches Elektronen-Synchrotron DESY,\\
Notkestra\ss e 85, \\ 
D-22607 Hamburg, Germany\\ 
E-mail: andreas.ringwald@desy.de}


\maketitle

\abstracts{
X-ray free electron lasers will be constructed in this decade,  
both at SLAC in the form of the so-called Linac Coherent Light 
Source as well as at DESY, where the so-called TESLA XFEL laboratory 
uses techniques developed for the design of the TeV energy superconducting electron-positron 
linear accelerator TESLA. 
Such X-ray lasers may allow also for 
high-field science applications by exploiting 
the possibility to focus their beams to a spot with a small radius, 
hopefully in the range of the laser wavelength. Along this route one  
obtains very large electric fields, much larger than those obtainable with any 
optical laser of the same power. We consider here 
the possibility of obtaining an electric field so high that electron-positron 
pairs are spontaneously produced in vacuum (Schwinger pair production) and 
review the prospects to verify this non-perturbative production 
mechanism for the first time in the laboratory.    
}

\section{Introduction}

Spontaneous particle creation from vacuum induced by an external field was first 
proposed in the context of 
$e^+e^-$ pair production in a static, spatially uniform electric 
field\cite{Sauter:1931} and is often 
referred to as the Schwinger\cite{Schwinger:1951nm} mechanism. It is one of the 
most intriguing non-linear phenomena in quantum field theory. Its consideration is  
theoretically important, since it requires one to go beyond 
perturbation theory, and its eventual experimental observation 
probes the theory in the domain of strong fields.  
Moreover, this mechanism has been applied to many problems in 
contemporary physics, ranging from black hole quantum 
evaporation\cite{Hawking:1974rv} and $e^+e^-$ creation in the vicinity of charged
black holes\cite{Damour:1974qv}, giving rise possibly to gamma ray bursts\cite{Preparata:1998rz},    
to particle production in hadronic 
collisions\cite{Casher:1979wy} 
and in the early universe\cite{Parker:1969au}, to 
mention only a few. One may consult the 
monographs\cite{Greiner:1985} for a review of further 
applications, concrete calculations and a detailed bibliography.

It is known since the early 1930's that in the background of a static, spatially 
uniform electric field the vacuum in quantum electrodynamics (QED) is unstable
and, in principle, sparks with spontaneous emission of $e^+e^-$ 
pairs\cite{Sauter:1931}.
However, a sizeable rate for spontaneous pair production requires 
extraordinary strong electric field strengths $\mathcal E$ of order or
above the critical value
\begin{equation}
{\mathcal E}_c \equiv \frac{m_e\, c^2}{e\, \lambdabar} 
= \frac{m_e^2\, c^3}{e\, \hbar}
\simeq 1.3\cdot 10^{18}\ {\rm V/m}\,. 
\label{schwinger-crit}
\end{equation}
Otherwise, for $\mathcal E\ll \mathcal E_c$, the work of the 
field on a unit charge $e$ over the Compton wavelength of the electron 
$\lambdabar =\hbar /(m_e c)$ is much smaller than the rest energy 
$2\,m_e c^2$ of the produced $e^+e^-$ pair, the
process can occur only via quantum tunneling, and its rate is  
exponentially suppressed, 
$\propto \exp [ -\pi\, \frac{{\mathcal E}_c}{\mathcal E} ]$.

Unfortunately, it seems inconceivable to produce macroscopic 
static fields 
with electric field strengths of the order of the critical 
field~(\ref{schwinger-crit}) in the laboratory. In view of this difficulty,    
in the early 1970's the question was raised whether intense optical lasers could be employed to 
study the Schwinger mechanism\cite{Bunkin:1970,Brezin:1970}. Yet, it was 
found that all available and conceivable optical lasers did not have enough 
power density to allow for a sizeable pair creation rate\cite{Bunkin:1970,%
Brezin:1970,Popov:1971,Troup:1972,Popov:1972b,Narozhnyi:1974,%
Mostepanenko:1974,Marinov:1977gq,Katz:1998,Dunne:1998ni,Fried:2001ga}. 
At about the same time, 
the thorough investigation of the question started whether the necessary 
superstrong fields around ${\mathcal E}_c$ can be generated microscopically and
transiently in the Coulomb field of colliding heavy ions with 
$Z_1+Z_2 > Z_c\approx 170$\cite{Zeldovich:1972}. At the present 
time, clear experimental signals for spontaneous positron creation in heavy 
ion collisions are still missing and could
only be expected from collisions with a prolonged lifetime\cite{Greiner:1998}.

Meanwhile, there are definite plans for the construction of X-ray 
free electron lasers (FEL), both 
at SLAC, where the so-called Linac Coherent Light Source\cite{Arthur:1998yq,Lindau:1999uw}
(LCLS) is under construction, 
as well as at DESY, where the so-called 
TESLA XFEL uses techniques developed for the design of the TeV energy superconducting $e^+e^-$  linear 
accelerator TESLA\cite{Brinkmann:1997nb,Materlik:1999uv,Materlik:2001qr}. 
Such X-ray lasers may possibly allow also for high-field science 
applications\cite{Melissinos:1998qn,Chen:1998,Chen:1999kp,Tajima:2000,Ringwald:2001cp}: 
One could make use of not only the high energy and transverse coherence of 
the X-ray beams, but also of 
the possibility to focus them to a spot with a small radius $\sigma$, 
hopefully in the range of the laser wavelength, 
$\sigma\,\gwig\, \lambda\simeq \mathcal{O}(0.1)$~nm. In this way one might 
obtain very large electric fields, 
\begin{eqnarray}
    \label{peak-electric-field}
    { \mathcal E} &= & \sqrt{
    \mu_0\,c\,
    \frac{P}{\pi \sigma^2} }
    \ = \ { 1.1\cdot 10^{17}}\ { \frac{\rm V}{\rm m}}\ 
    \left( \frac{P}{1\ {\rm TW}}\right)^{1/2}\,
    \left( \frac{0.1\ {\rm nm}}{\sigma}\right)\,,
\end{eqnarray}
much larger than those obtainable with any optical laser of the same peak power $P$.  
Thus, X-ray FELs may be employed possibly as vacuum boilers\cite{Chen:1998}.

In this contribution, I will review recent work on spontaneous $e^+e^-$ pair production 
at the focus of future X-ray FELs\cite{Ringwald:2001ib,Alkofer:2001ik,Roberts:2002py,Popov:ak} 
and discuss the prospects to verify this non-perturbative production 
mechanism for the first time in the laboratory. 

\section{X-Ray Free Electron Lasers}

Let us start by briefly reviewing 
the principle of X-ray free electron lasers.

\begin{figure}[t]
\centerline{
\epsfxsize=10.5cm\epsfbox{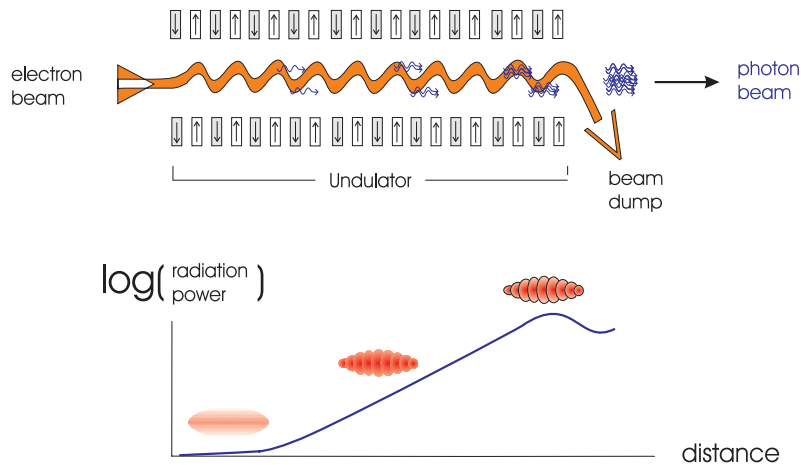}
                   }   
\caption[]{Principle of a single-pass  X-ray free electron laser in the self amplified spontaneous 
                      emission  mode\cite{Materlik:2001qr}.\label{fig:xfel_princ}}
\end{figure}

Conventional lasers yield radiation typically in the optical band. The reason is 
that in these devices the gain comes from stimulated emission from electrons
bound to atoms, either in a crystal, liquid dye, or a gas.
The amplification medium of free electron lasers\cite{Madey:1971}, 
on the other hand, is {\em free}, i.e. unbounded, electrons
in bunches accelerated to relativistic velocities with a characteristic longitudinal
charge density modulation (cf. Fig.~\ref{fig:xfel_princ}). 

\begin{figure}[t]
\begin{center}
\epsfig{file=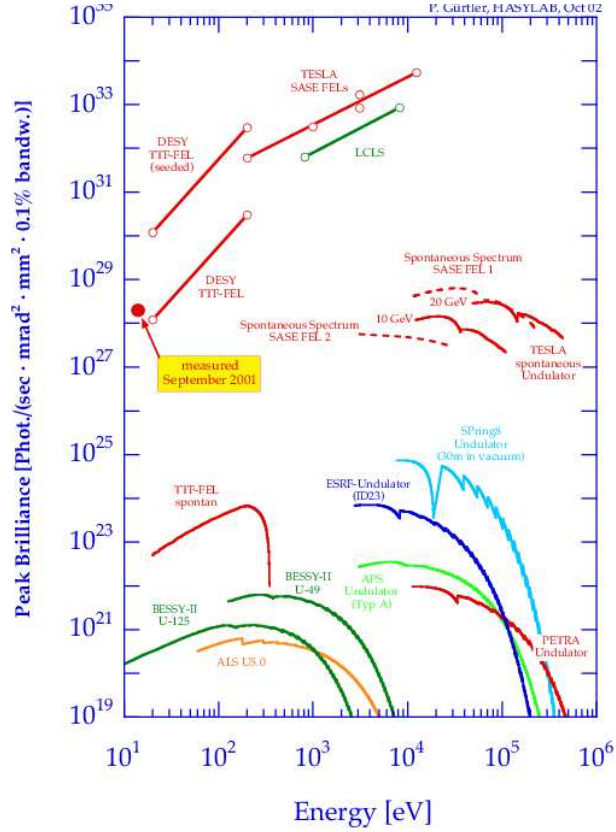,bbllx=29pt,%
        bblly=250pt,bburx=390pt,bbury=748pt,width=8cm,clip=}
\caption[]{Spectral peak brilliance of X-ray FELs and undulators for 
                      spontaneous radiation at TESLA, together with that of third generation
                     synchrotron radiation sources\cite{Materlik:2001qr}. 
                       For comparison, the spontaneous 
                      spectrum of an X-ray FEL undulator is shown.
\label{fig:xfel_spect}}
\end{center}
\end{figure}

\begin{figure}[t]
\begin{center}
\epsfig{file=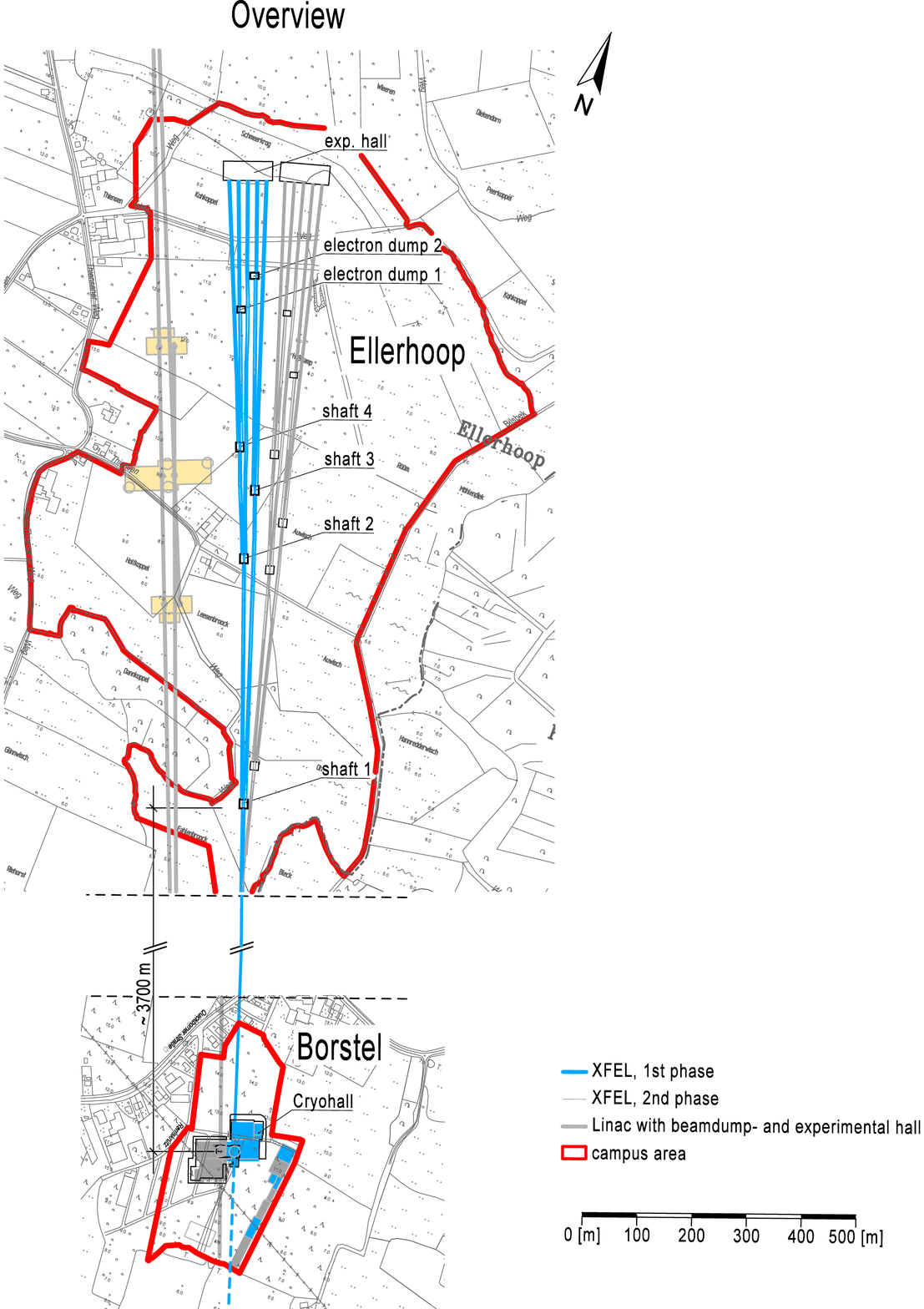,angle=180,width=8cm,clip=}
\caption[]{The TESLA XFEL campus North-West of the DESY laboratory\cite{Materlik:2001qr}, whose
commissioning is expected in 2010. 
The XFEL electron beam is accelerated by a dedicated $20$~GeV linear accelerator (linac) starting
at a supply hall $\approx 4$~km south of the XFEL laboratory. The XFEL linac tunnel runs under
a small angle of $2^{\rm o}$ with respect to the tunnel of the future TESLA linac, which
is shown in grey color. 
\hfill
\label{fig:xfel_camp}}
\end{center}
\end{figure}

The basic principle of a single-pass free electron laser operating in the self amplified
spontaneous emission (SASE) mode\cite{Kondratenko:1980} is as follows.
It functions by passing an electron beam pulse of energy $E_e$ of small 
cross section and high peak current through an undulator -- a long periodic magnetic structure   
(cf. Fig.~\ref{fig:xfel_princ}). 
The interaction of the emitted synchrotron radiation, with opening angle
\begin{equation}
     1/\gamma = m_e c^2/E_e 
= 2\cdot 10^{-5}\ \left(  25\ {\rm GeV}/E_e
\right) \,,
\end{equation}
where $m_e$ is the electron mass, 
with the electron beam pulse within the undulator leads to the buildup of a
longitudinal charge density modulation (micro bunching), if a resonance condition,
\begin{equation}  
     \lambda = \frac{\lambda_{\rm U}}{2\gamma^2}
     \left( 1 + \frac{K^2_{\rm U}}{2}\right)
= 0.3\ {\rm nm}\ \left( \frac{\lambda_{\rm U}}{1\,{\rm m}}\right) 
\left( \frac{1/\gamma}{2\cdot 10^{-5}}\right)^2\,
\left( \frac{ 1 + K^2_{\rm U}/2 }{3/2}\right)
\,, 
\end{equation}
is met. Here, $\lambda$ is the wavelength of the emitted radiation, 
$\lambda_{\rm U}$ is the length of the magnetic period of the undulator, and
$K_{\rm U}$ is the undulator parameter,
\begin{equation}
K_{\rm U} 
= \frac{e \lambda_{\rm U} B_{\rm U} }{
                2\pi m_e c}\,, 
\end{equation}
which gives the ratio between the average deflection angle of the electrons in the 
undulator magnetic field $B_{\rm U}$ from the
forward direction and the 
typical opening cone of the synchrotron radiation. The undulator parameter should be
of order one on resonance. 
The electrons in the developing micro bunches eventually radiate
coherently -- the gain in radiation power $P$, 
\begin{equation}
P\propto e^2\,{N_e^2}\,B_{\rm U}^2\,\gamma^2\,,
\end{equation}
over the one from incoherent spontaneous synchrotron radiation ($P\propto N_e$) 
being proportional to the number $N_e\geq 10^9$ of electrons in a bunch 
(cf. Fig.~\ref{fig:xfel_spect}) --  
and the number of emitted photons grows exponentially
until saturation is reached.
The radiation has a high power, short pulse length, narrow bandwidth, is fully polarized, 
transversely coherent, and has a tunable wavelength. 

The concept of using a high energy electron linear accelerator for building an 
X-ray FEL was first proposed for the Stanford Linear Accelerator\cite{Arthur:1998yq}. 
The LCLS at SLAC is expected to provide the first X-ray laser beams in 2008.   
The feasibility
of a single-pass FEL operating in the SASE mode has been demonstrated 
recently  down to a wavelength of 80 nm using electron bunches of high charge density and low
emittance from the linear accelerator at the TESLA test facility (TTF) at DESY\cite{Andruszkow:2000it}  
(cf. Fig.~\ref{fig:xfel_spect}). 
Some characteristics of the radiation from the planned 
X-ray FELs at the TESLA XFEL laboratory\cite{Materlik:2001qr} (cf. Fig.~\ref{fig:xfel_camp}), 
whose commissioning is expected in 2010, are listed in Table~\ref{tab:xfel_par}. 

\begin{table}[ph]
\tbl{Properties of X-ray FELs at the TESLA XFEL laboratory.\hfill\vspace*{1pt}}
{\footnotesize
\begin{tabular}{|lc||c|c|c|}\hline 
 & unit & SASE 1 & SASE 3 & SASE 5\\\hline
wavelength  & nm & $0.1\div 0.5$ & $0.1\div 0.24$ & $0.4\div 5.8$ \\
bandwidth (FWHM) & \% &0.08&0.08&$0.29\div 0.7$\\
peak power & GW & 37 & 22 & $110\div 200$\\ 
average  power & W & 210 & 125 & $610\div 1100$\\
photon beam size (rms)  & $\mu$m & 43 & 53 & $25\div 38$\\
peak power density & W/m$^2$ & $6\cdot 10^{18}$ & $3\cdot 10^{18}$ & $6\cdot 10^{19}$
\\\hline
\end{tabular}
\label{tab:xfel_par} 
       }
\vspace*{-13pt}
\end{table}

\section{Semi-classical Rate Estimates}

We now turn to the main subject of our contribution, namely the spontaneous pair
production at the focus of future X-ray FELs. 
We will elaborate in this section on a simplified approximation concerning the electromagnetic field 
of the laser radiation which retains the main features of the general case but 
nevertheless allows to obtain final expressions for the pair production rate in 
closed form. This should be sufficient for an order-of-magnitude estimate of critical 
parameters to be aimed at to get an observable effect.

\begin{table}[t] 
\tbl{ 
Laser parameters and derived quantities relevant for 
estimates of the rate of spontaneous $e^+e^-$ pair production. 
The column labeled ``Optical'' lists parameters which are typical for
a petawatt-class (1 PW = $10^{15}$ W) optical laser, 
focused to the diffraction limit, 
$\sigma = \lambda$. The column labeled ``Design'' displays design 
parameters of the planned X-ray FELs at DESY 
(Table~\ref{tab:xfel_par}). Similar values apply for 
LCLS.
The column labeled ``Focus: Available'' shows typical values which can
be achieved with present day methods of X-ray 
focusing: It assumes that the X-ray FEL 
X-ray beam can be focused to a rms spot radius of $\sigma \simeq 21$ nm with 
an energy extraction efficiency of 1 \%. The column 
labeled ``Focus: Goal'' 
shows parameters which are theoretically possible by increasing the energy 
extraction of LCLS (by the tapered undulator technique) and by a yet 
unspecified method of diffraction-limited focusing of X-rays\vspace*{1pt}.} 
{\footnotesize
\begin{tabular}{|c|c||c|c|c|}\hline 
\multicolumn{5}{|c|}{{\bf Laser Parameters}}\\\hline
  & {\bf Optical}  & \multicolumn{3}{c|}{\bf X-ray FEL}
\\\hline
  & Focus: & Design & Focus: & 
Focus: \\  
    & Diffraction limit &  & Available & 
Goal  
\\\hline 
 $\lambda$ & 1 $\mu$m & 0.4 nm & 0.4 nm & 0.15 nm \\
 $\hbar\,\omega = \frac{hc}{\lambda}$ & 1.2 eV & 3.1 keV 
& 3.1 keV 
& 8.3 keV\\
$P$ & 1 PW & 110 GW  & 1.1 GW &  5 TW                \\
$\sigma$ & 1 $\mu$m & 26 $\mu$m & 21 nm  & 0.15 nm\\
$\triangle t $ & 500 fs $\div$ 20 ps
& 0.04 fs
& 0.04 fs
& 0.08 ps               \\\hline
\multicolumn{5}{|c|}{{\bf Derived Quantities}}\\\hline
 & & & & \\[0.5ex]
$S=\frac{P}{\pi \sigma^2 }$& $3\times 10^{26}$ 
$\frac{\rm W}{{\rm m}^2}$
& $5\times 10^{19}$ $\frac{\rm W}{{\rm m}^2}$ & 
$8\times 10^{23}$ $\frac{\rm W}{{\rm m}^2}$ & $7\times 10^{31}$ 
$\frac{\rm W}{{\rm m}^2}$
\\[1ex]
$\mathcal E =\sqrt{\mu_0\,c\, S}$
& $4\times 10^{14}$ $\frac{\rm V}{\rm m}$ & 
$1\times 10^{11}$ $\frac{\rm V}{\rm m}$ &
$2\times 10^{13}$ $\frac{\rm V}{\rm m}$ & 
$2\times 10^{17}$ $\frac{\rm V}{\rm m}$
\\[1ex]
$\mathcal E/{\mathcal E}_c$  &$3\times 10^{-4}$ & $1\times 10^{-7}$   
& $1\times 10^{-5}$  & 0.1     
\\
$\frac{\hbar\omega}{m_e c^2}$ & 
 $2\times 10^{-6}$ & $0.006$ &
$0.006$ & $0.02$ \\
$\eta = 
\frac{\hbar \omega}{e\, \mathcal E \lambdabar}$ 
& $9\times 10^{-3}$ & $6\times 10^{4}$ &
$5\times 10^{2}$ & 0.1 \\
\hline
\end{tabular}
\label{parameters}
}
\vspace*{-13pt}
\end{table}

It is well known that no pairs are
produced in the background of a light-like static, spatially uniform 
electromagnetic field\cite{Schwinger:1951nm}, 
characterized invariantly by
\begin{eqnarray}
\label{F}
\mathcal F &\equiv & \frac{1}{4}\,F_{\mu\nu}F^{\mu\nu}\equiv 
-\frac{1}{2}\,(\mathbf E^2 -c^2\mathbf B^2)=0\,,
\\[1ex] 
\label{G}
\mathcal G &\equiv & \frac{1}{4}\,F_{\mu\nu}\tilde F^{\mu\nu}\equiv 
c\,\mathbf E\cdot\mathbf B =0\,,
\end{eqnarray}
where $F^{\mu\nu}$ is the electromagnetic field strength tensor and 
$\tilde F^{\mu\nu} =(1/2)\,\epsilon^{\mu\nu\alpha\beta}F_{\alpha\beta}$ its 
dual. It has been argued that 
fields produced by focusing laser beams are very close to
such a light-like electromagnetic field, leading to an essential suppression 
of pair creation\cite{Troup:1972}. Yet, in a focused wave there are regions near 
the focus where 
$\mathcal F<0$ and pair production is 
possible\cite{Bunkin:1970,Melissinos:1998qn}.
For other fields, $\mathcal F$ and $\mathcal G$ do not vanish, and pair
production becomes possible, unless $\mathcal G =0$, $\mathcal F>0$, 
corresponding to a pure magnetic field in an appropriate coordinate 
system\cite{Schwinger:1951nm}. 
In particular, one expects pair creation in the background of 
a spatially uniform 
electric field oscillating with a frequency $\omega$, say
\begin{equation}
{\mathbf E}(t) =(0,0,{\mathcal E}\cos (\omega t))\,,
\hspace{6ex}
{\mathbf B}(t) =(0,0,0)\,,
\label{electric-type}
\end{equation}
which has $\mathcal G =0$, $\mathcal F<0$. As emphasized in 
Refs.\cite{Popov:1972b,Mostepanenko:1974,Marinov:1977gq,Chen:1998}, 
such a field may be created in an antinode of the standing wave 
produced by a superposition of two coherent laser beams with wavelength 
\begin{equation}
\lambda = \frac{2\pi c}{\omega}\,,
\end{equation}
and, indeed, it may be considered as 
spatially uniform at distances much less than the wavelength.

Thus, for definiteness, we assume that every X-ray laser pulse is 
split into two equal parts and recombined to form a standing wave with
locations where the electromagnetic field has the 
form~(\ref{electric-type}) and 
where the peak electric field is given by Eq.~(\ref{peak-electric-field}). 
Alternatively, one may consider pair creation in the overlap region of two lasers, whose
beams make a fixed angle to each other\cite{Fried:2001ga}. 
Furthermore, we assume that 
the field amplitude $\mathcal E$ is much smaller than the critical field,
and the photon energy is much smaller than the rest energy of the electron, 
\begin{equation}
{\mathcal E}\ll {\mathcal E}_c\,,\hspace{6ex}
\hbar\omega\ll m_e c^2\,;
\label{conditions}
\end{equation}
conditions which are well satisfied at realistic X-ray lasers 
(cf. Table~\ref{parameters}). 
Under these conditions, it is possible to compute the rate of 
$e^+e^-$ pair production in a semi-classical manner, using 
generalized WKB or imaginary-time (instanton) 
methods\cite{Brezin:1970,Popov:1971,Dunne:1998ni,Fried:2001ga,Kim:2000un}. 
Here, the ratio $\eta$ of the energy of the laser photons over 
the work of the field on a unit charge $e$ over the Compton wavelength of the 
electron,
\begin{equation}
\label{gamma}
\eta = \frac{\hbar\omega}{e {\mathcal E} \lambdabar }
=
\frac{\hbar\,\omega}{m_e c^2}\,\frac{{\mathcal E}_c}{\mathcal E} =
\frac{m_e c\,\omega}{e\,\mathcal E}\,,
\end{equation}
plays the role of an adiabaticity parameter.  
Indeed, the probability that an $e^+e^-$ pair is 
produced per unit time and unit volume,    
\begin{eqnarray}
w = 
\frac{{\rm d}\,n_{e^+e^-}}{{\rm d}^3x\,{\rm d}t}    
\,,
\end{eqnarray}
depends on the laser frequency only through the adiabaticity parameter $\eta$
and reads, in the limiting cases of small and large $\eta$, 
as follows\cite{Ringwald:2001ib,Popov:ak} 
\begin{eqnarray}
\label{w_popov_lim}
\lefteqn{
w \simeq  \frac{c}{4\,\pi^3 \lambdabar^4}\,\times}
\\[1ex] \nonumber && 
\times\,\left\{
\begin{array}{lcr}
\frac{\sqrt{2}}{\pi} 
 \left( 
\frac{\mathcal E}{{\mathcal E}_c}
\right)^{\frac{5}{2}}\,
\exp \left[ 
-\pi\, \frac{{\mathcal E}_c}{\mathcal E} 
\left( 1-\frac{1}{8}\eta^2+\mathcal O(\eta^4)\right)
\right]\,,
&:& \eta\ll 1\,,\\[2ex]
\sqrt{\frac{\pi}{2}} 
\left( 
\frac{\hbar\,\omega}{m_e c^2}
\right)^{\frac{5}{2}}
\sum_{n>2\frac{m_ec^2}{\hbar \omega}}
\left( \frac{{\rm e}}{4\eta}\right)^{2n}
{\rm e}^{ 
-2\left( n-2\frac{m_ec^2}{\hbar \omega}\right)}\times
&&
\\[1.5ex] 
\times\, {\rm Erfi} \left( 
\sqrt{2\left( n-2\frac{m_ec^2}{\hbar \omega}\right)}\right)
&:& \eta\gg 1\,,
\end{array}
\right.
\end{eqnarray}
where $\rm Erfi$ is the imaginary error function. 
This result agrees in the adiabatic high-field, low-frequency limit, $\eta\ll 1$, 
with the non-perturbative Schwinger result\cite{Schwinger:1951nm} for a static, spatially uniform field, 
if a proper average over an oscillation period is made.  
In the non-adiabatic low-field, high-frequency limit, $\eta\gg 1$, on the other hand, 
it resembles a perturbative result: it corresponds to  
the $\geq n$-th order perturbation theory, $n$ being the minimum number of quanta
required to create an $e^+e^-$ pair: $n\gwig 2\, m_e c^2/(\hbar\omega)\gg 1$.

\begin{figure}[t]
\begin{center}
\epsfig{figure=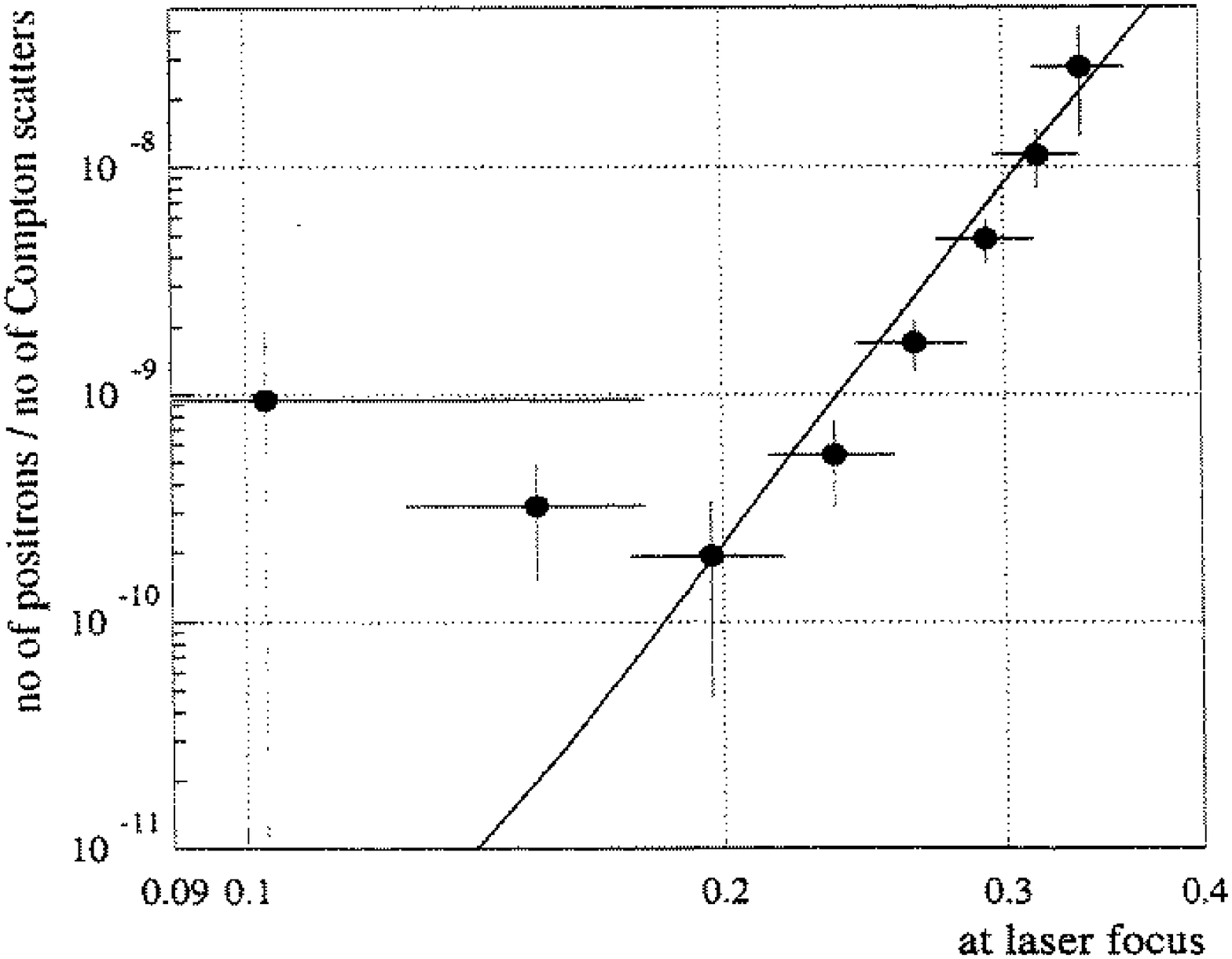,width=9cm,clip=}
\put(-120,0){\Large $\eta^{-1}$}
\caption[...]{The positron rate per laser shot as a function of the 
inverse of the adiabaticity parameter, $\eta^{-1}$, as measured 
by the SLAC experiment E-144\cite{Burke:1997ew}. The line is a power law fit
to the data which gives $R_{e^+}\propto \eta^{-2n}$, with 
$n=5.1\pm 0.2\,({\rm stat})^{+0.5}_{-0.8}\,({\rm syst})$. 
\hfill
\label{fig:e144}}
\end{center}
\end{figure}

At this point it seems appropriate to discuss the question whether -- 
as argued in Ref.\cite{Melissinos:1998qn} --  
the non-perturbative Schwinger pair creation mechanism has already 
been demonstrated by the SLAC experiment E-144\cite{Burke:1997ew}. This experiment studied 
positron production in the collision of $46.6$ GeV/c electrons with
terawatt optical ($\lambda = 527$ $\mu$m) laser pulses.  
In the rest frame of the incident electrons, an electrical field strength 
of about $38$~\% of the critical field~(\ref{schwinger-crit}), 
${\mathcal E} \simeq 5\cdot 10^{17}\ {\rm V/m}$, was reached.  
The values of the adiabaticity parameter $\eta$ probed were therefore
in the range $\eta\simeq 3\div 10$ (cf. Fig.~\ref{fig:e144}), i.e. in the 
non-adiabatic, perturbative multi-photon regime. Correspondingly, in Refs.\cite{Burke:1997ew,Bamber:1999zt}
the data were convincingly interpreted in terms of multi-photon light-by-light scattering.
Indeed, the observed positron production rate scales as $R_{e^+}\propto \eta^{-10}$ (cf. Fig.~\ref{fig:e144}). 
This is in good agreement with the fact that the rate of perturbative multi-photon reactions 
involving $n$ laser photons is proportional to $\eta^{-2n}$ for $\eta\gg 1$, Eq.~(\ref{w_popov_lim}), 
and with the kinematic requirement that five photons are needed to produce a pair 
near threshold.  

For an X-ray laser ($\hbar\omega\simeq 1\div 10$~keV),  
the adiabatic, non-perturbative, strong field regime, $\eta\,\lwig\, 1$, 
starts to apply for 
$\mathcal E\,\gwig\, \hbar\omega\, 
{\mathcal E}_c/(m_ec^2)\sim 10^{15\div 16}$ V/m 
(cf. Eq.~(\ref{gamma})). 
An inspection of the rate~(\ref{w_popov_lim}) leads then to the conclusion that   
one needs an electric field of about $0.1\,{\mathcal E}_c\sim 10^{17}$ V/m 
in order to get an appreciable amount of spontaneously produced $e^+ e^-$ pairs\cite{Ringwald:2001ib}. 
To this end one needs either a terawatt X-ray or a tens of exawatt optical laser.

In Table~\ref{parameters} we have summarized the relevant parameters for the 
planned X-ray FELs\cite{Ringwald:2001ib}. We conclude that 
the power densities and electric fields which can be reached with presently
available technique (column labeled ``Focus: Available'' in Table~\ref{parameters})
are far too small for a sizeable effect. On the other hand, if the  
energy extraction can be improved considerably, such that 
the peak power of the planned X-ray FELs can be increased to the terawatt region,
and if X-ray optics can be improved\cite{Hastings:2000} to approach the diffraction limit of focusing,
leading to a spot size in the 0.1 nanometer range, then there is ample
room (c.\,f. column labeled ``Focus: Goal'' in Table~\ref{parameters}) for an 
investigation of the Schwinger pair production mechanism at 
X-ray FELs. At the moment it is hard to predict whether this goal will 
be reached before the commissioning of exawatt-zettawatt optical lasers\cite{Tajima:2002}.  

\begin{figure}[t]
\begin{center}
\epsfig{figure=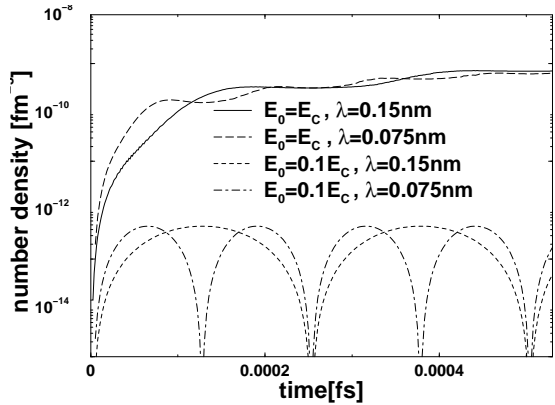,angle=-90,width=7.2cm,clip=}
\caption[...]{Time evolution of the number density of produced
$e^+e^-$ pairs at the focus of an X-ray laser\cite{Alkofer:2001ik}. 
In strong fields, particles accumulate, leading to 
the almost complete occupation of available momentum 
states. In weak fields, repeated cycles of particle creation and
annihilation occur in tune with the laser frequency.
\hfill
\label{fig:numb_dens}}
\end{center}
\end{figure}

\section{Quantum Kinetic Studies}

\begin{figure}[t]
\begin{center}
\epsfig{file=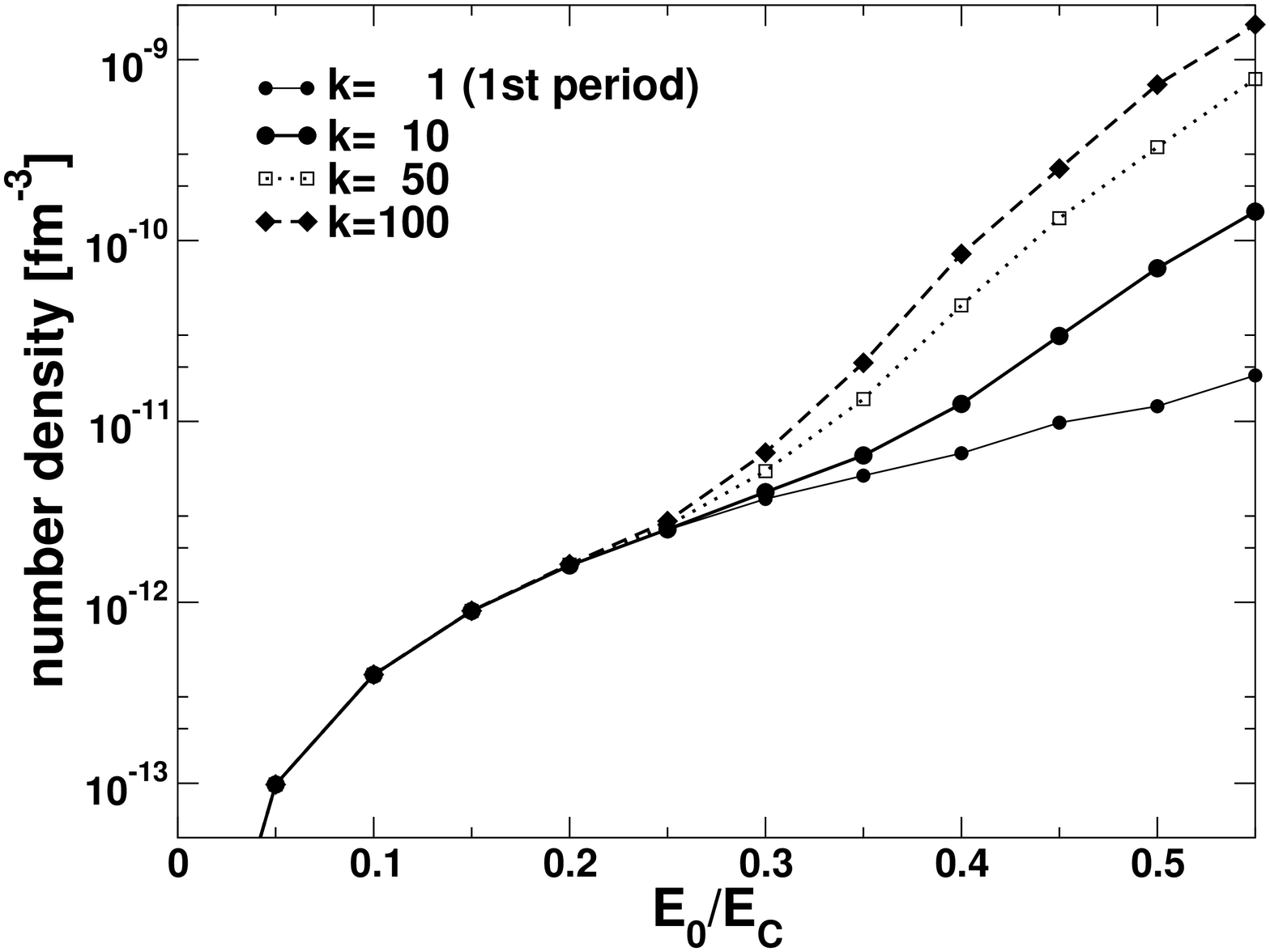,width=5.65cm,clip=}
\hfill
\epsfig{file=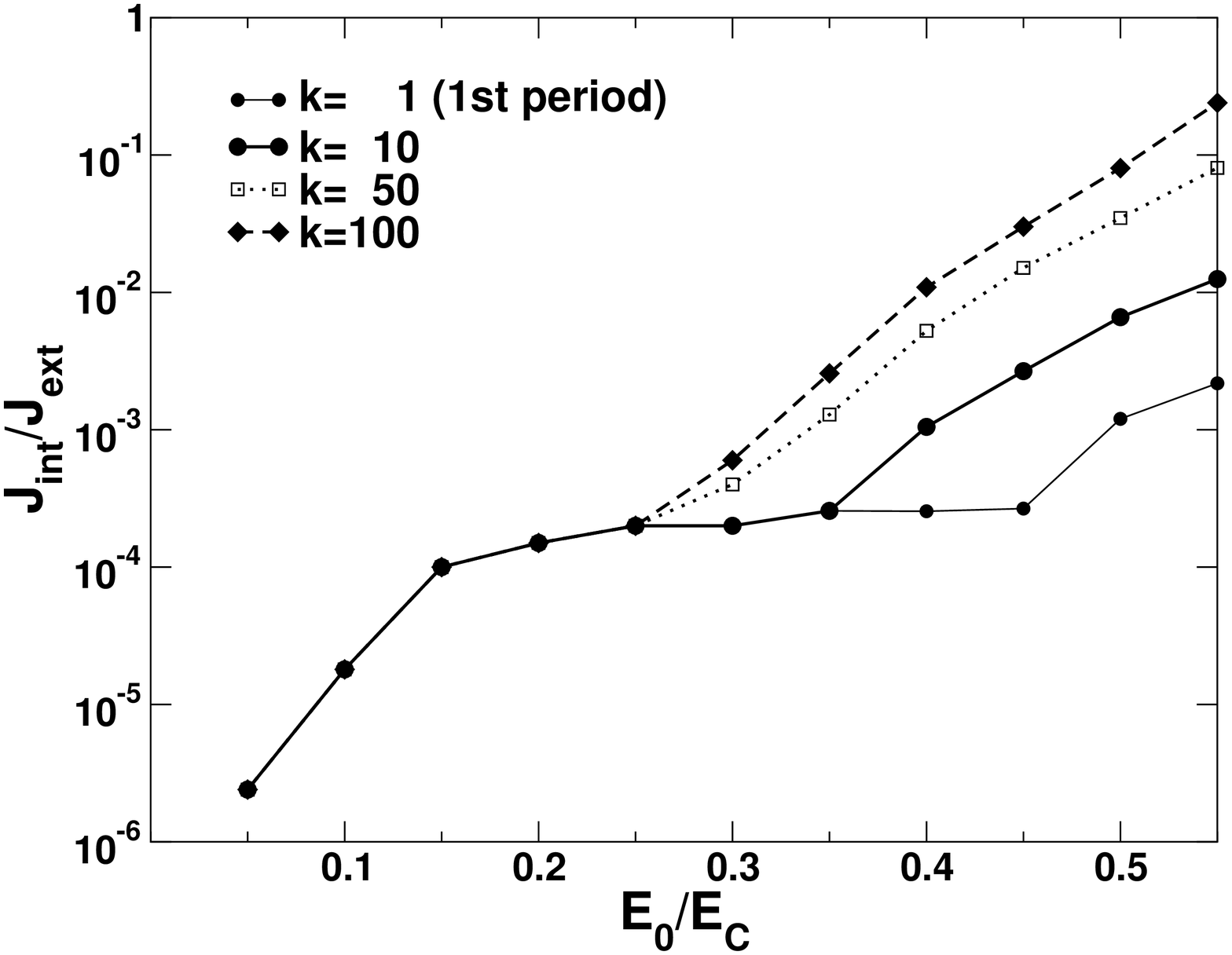,width=5.65cm,clip=}
\caption[...]{{\em Left:} Peak particle number density versus laser
field strength\cite{Roberts:2002py}. The qualitative change at $E_0\approx 0.25\,E_c$
marks the onset of particle accumulation. 
{\em Right:} Internal to external peak current ratio\cite{Roberts:2002py}: 
field-current feedback becomes
important for $E_0\gwig 0.25\,E_c$.  
\hfill
\label{fig:numb_accum}}
\end{center}
\end{figure}

More information about the details of the Schwinger mechanism accessible at 
the focus of an X-ray laser can be 
obtained via approaches based on quantum kinetics. 
In Refs.\cite{Alkofer:2001ik,Roberts:2002py}, quantum Vlasov equations, 
derived within a mean-field treatment of QED\cite{Smolyansky:1997fc}, were employed to obtain
a description of the time evolution of the momentum distribution function
for the particles produced via vacuum decay in the background of a spatially
uniform external electric field of the form (\ref{electric-type}).
It was found that -- for realistic laser parameters (cf. Table~\ref{parameters}) -- 
pair production will occur in cycles that proceed
in tune with the laser frequency (cf. Fig.~\ref{fig:numb_dens}). 
The peak density of produced pairs, however, is frequency independent, with 
the consequence that several hundred pairs could be produced per laser period, 
in accord with the Schwinger rate. For even higher peak electric fields, 
${\mathcal E}\gwig 0.25\,{\mathcal E}_c$ -- possibly achievable at a $9$~TW X-ray 
FEL (cf. Table~~\ref{parameters}) -- particle accumulation and 
the consequent formation of a plasma of spontaneously produced pairs 
is predicted\cite{Roberts:2002py} (cf. Fig.~\ref{fig:numb_accum}). 
The evolution of the particle number in the plasma will exhibit then non-Markovian
aspects, and the plasma's internal currents will generate
an electric field whose interference with that of the laser leads to plasma 
oscillations\cite{Roberts:2002py}. This feature persists even if -- in distinction to
Refs.\cite{Alkofer:2001ik,Roberts:2002py} -- one takes into account collision terms in the 
quantum Vlasov equations\cite{Ruffini:2003cr}. 

\section{Conclusions}

We have considered the possibility to study non-perturbative spontaneous 
$e^+e^-$ pair creation from vacuum for the first time in the laboratory.  
We have seen that for this application still some improvement in X-ray FEL 
technology over the presently considered design parameters is necessary. 
Intensive development in technical areas, particularly in that of X-ray
optics, will be needed in order to achieve the required ultra-high 
power densities.
It should be pointed out, however, that even though progress to achieve such 
a demanding goal is rather slow and laborious, the rewards that may be gained in 
this unique regime are so extraordinary that looking into TESLA XFEL's or 
LCLS's extension to this regime merits serious considerations.
No doubt, there will be unprecedented opportunities to use these intense X-rays in order
to explore some issues of fundamental physics that have eluded
man's probing so far.

\end{document}